\documentclass[pra,superscriptaddress, twocolumn]{revtex4-2}
\usepackage[dvipsnames]{xcolor} 
\definecolor{IndustrialBlue}{RGB}{29,88,167}

\usepackage{tikz}
\usetikzlibrary{arrows.meta}
\usepackage{soul}     
\soulregister\cite7   
\soulregister\ref7    
  
\usetikzlibrary{positioning}

\usepackage{amsmath}
\usepackage{physics}
\usepackage{graphicx}
\usepackage{dcolumn}
\usepackage{bm}
\usepackage[colorlinks=true,urlcolor=IndustrialBlue,citecolor=IndustrialBlue,linkcolor=IndustrialBlue]{hyperref}
\usepackage{comment}

\usepackage{appendix}

\usepackage{placeins}

\definecolor{red}{RGB}{255,87,87}
\definecolor{blue}{RGB}{82,113,255} 

\begin{document}

\preprint{APS/123-QED}

\title{High-Distance Error-Correcting Codes for Fermion-to-Qubit Mappings in 2D and 3D}
\author{Ruby Wei}
\email{Ruby.Wei@colorado.edu}
\affiliation{JILA, NIST and University of Colorado, Boulder, Colorado 80309, USA}
\affiliation{Department of Physics, University of Colorado, Boulder, Colorado 80309, USA}
\author{Aqua Chung}
\affiliation{JILA, NIST and University of Colorado, Boulder, Colorado 80309, USA}
\affiliation{Department of Physics, University of Colorado, Boulder, Colorado 80309, USA}
\author{Luke Coffman}
\affiliation{JILA, NIST and University of Colorado, Boulder, Colorado 80309, USA}
\affiliation{Department of Physics, University of Colorado, Boulder, Colorado 80309, USA}
\author{Su-Kuan Chu}
\email{Su-Kuan.Chu@colorado.edu}
\affiliation{JILA, NIST and University of Colorado, Boulder, Colorado 80309, USA}
\affiliation{Department of Physics, University of Colorado, Boulder, Colorado 80309, USA}
\author{Xun Gao}
\affiliation{JILA, NIST and University of Colorado, Boulder, Colorado 80309, USA}
\affiliation{Department of Physics, University of Colorado, Boulder, Colorado 80309, USA}

\begin{abstract}
Quantum simulation of fermionic systems is a leading application of quantum computers. One promising approach is to represent fermions with qubits via fermion-to-qubit mappings. In this work, we present high-distance fermion-to-qubit stabilizer codes for simulating 2D and 3D fermionic systems. These codes achieve arbitrarily large code distances while keeping stabilizer weights constant. They also preserve locality by mapping local fermionic operators to local qubit operators at any fixed distance. Notably, our 3D construction is the first to simultaneously achieve high distance, constant stabilizer weights, and locality preservation. Our construction is based on concatenating a small-distance 2D or 3D fermion-to-qubit code with a high-distance fermionic color code. Together, these features provide a robust and scalable pathway to quantum simulation of fermionic systems.
\end{abstract}
\maketitle

\section{Introduction} \label{sec:overview}
The development of quantum computers is largely motivated by the ability to simulate physical systems~\cite{feynman_simulating_1982}, with fermionic systems being particularly appealing due to applications in quantum chemistry, condensed matter, and high-energy physics~\cite{cao2019quantum, mcardle_quantum_2020}. As the majority of quantum computing architectures employ qubits, it is crucial to find good fermion-to-qubit mappings where Hamiltonians of fermionic systems (fermionic operators) are mapped to Pauli operators in order to facilitate simulation.

The earliest fermion-to-qubit mapping is the famous Jordan-Wigner transformation (JWT)~\cite{jordan_uber_1928}, which maps fermionic creation and annihilation operators to strings of Pauli operators. In 1D, the JWT preserves locality by mapping local fermionic operators to local Pauli operators. In higher dimensions, however, it maps local fermionic operators to nonlocal Pauli strings whose weight scales with system size, imposing limits on scalable implementations. Furthermore, it lacks robustness to noise: qubit errors are always mapped to fermionic errors, making it highly error-prone in practice.

For a fermion-to-qubit mapping to be locality-preserving, it must map fermionic operators supported on neighboring sites to qubit operators supported on neighboring qubits, in terms of their respective spatial geometries. Since many Hamiltonians of interest from condensed matter physics are local (for example, the Hubbard model~\cite{fradkin_field_2013}), locality-preserving maps reduce the overhead in simulation and physical implementation~\cite{nys2023quantum}. However, because the encoding must satisfy anticommutation relations, to have locality at the operator level, long-range entanglement must be present at the state level in dimensions higher than 1D~\cite{bausch2020mitigatingerrorslocalfermionic,kobayashi2025generalizedstatisticslattice,Li2025,guaita_locality_2025}.

Beyond preserving locality, robustness to noise is equally crucial for practical quantum simulation, motivating the introduction of redundancy to protect the information of fermions. This lends itself naturally to the stabilizer formalism, where fermion-to-qubit mappings can be interpreted as quantum error-correcting codes~\cite{verstraete2005mapping,chen_exact_2018, chen_bosonization_2019,  jiang2019majorana,setia2019superfast,chen2020,derby2021compact,chien2022optimizing, chen2023equivalence, landahl2023logicalfermionsfaulttolerantquantum, chen_error-correcting_2024,Algaba_2024,iv2024low,algaba_fermion-qubit_2025}. Importantly, for experimental feasibility, it is desirable to have low-weight stabilizers, in that the weights scale independently of the code distance. In total, finding robust fermion-to-qubit encodings that can resist errors on a large number of qubits while still being local in multiple dimensions is a significant step in realizing quantum simulation.

While many fermion-to-qubit mappings have been developed, they often lack error-correcting properties or have a small code distance~\cite{Wosiek:1981mn,verstraete2005mapping,ball2005fermions,Kitaev_2006,farrelly2014causal,whitfield2016local,chen_exact_2018, chen_bosonization_2019, steudtner2019quantum, jiang2019majorana,chen2020,Bochniak2020mko,derby2021compact,po2021symmetric,chien2022optimizing, li2022,chen2023equivalence,chien2023simulating,chiew2023discovering,chen_error-correcting_2024,o2024ultrafast,Algaba_2024, harrison2024sierpinskitrianglefermiontoqubittransform,chen2025neural}. Recent efforts have improved the code distance by exploiting excitations~\cite{chen_exact_2018,chen_error-correcting_2024} or twist defects~\cite{landahl2023logicalfermionsfaulttolerantquantum,algaba_fermion-qubit_2025} in topological codes. However, finding locality-preserving codes with larger distances either requires an expensive computational search ~\cite{chen_error-correcting_2024, iv2024low} or is limited to 1D or 2D systems ~\cite{landahl2023logicalfermionsfaulttolerantquantum,chen_error-correcting_2024, iv2024low, algaba_fermion-qubit_2025}. Since many interesting physical problems in the real world involve simulating fermions in 3D, such as quantum chemistry~\cite{mcardle_quantum_2020}, condensed matter~\cite{Xu_2013, armitage2018weyl}~(including superconductivity~\cite{bansil2005}), and quantum chromodynamics~\cite{ciavarella2023quantum}, finding locality-preserving 3D encodings with arbitrary code distances remains a crucial step towards more efficient quantum simulation.

\begin{figure}
    \centering
    \includegraphics[width=\linewidth]{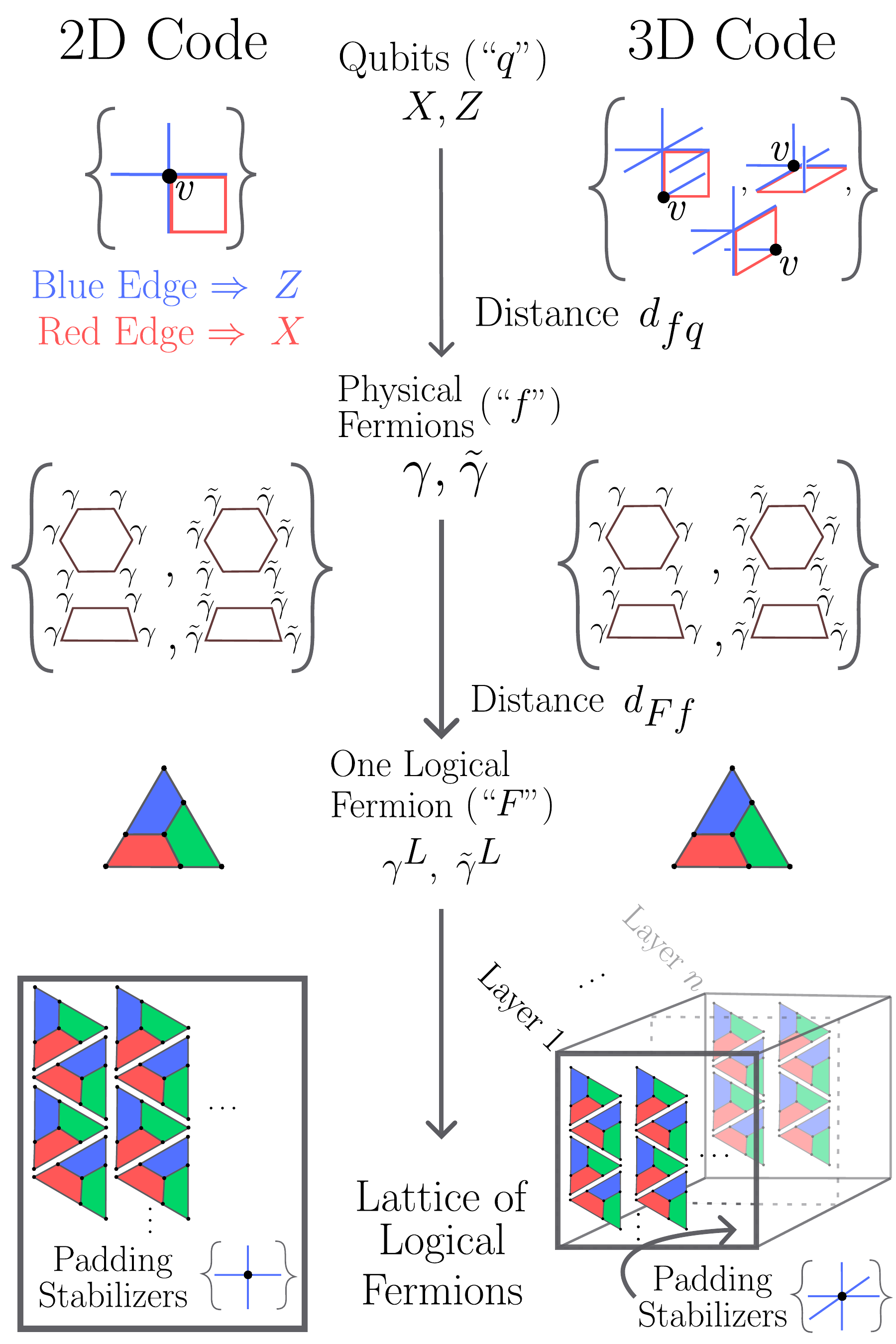}
     \caption{Overview of our 2D and 3D fermion-to-qubit codes. From the bottom of the figure to the top, we see that logical fermions on a lattice are built from physical fermions arranged into 2D fermionic color codes, and each ``physical" fermion emerges from the excitations of topologically ordered systems based on qubits. Stabilizers are constructed to obtain the desired algebra and codespace. Our code, with the overall distance labeled as $d_{Fq}$, is a concatenation of a fermionic code (with a large distance $d_{Ff}$) and a fermion-to-qubit code (with a small distance $d_{fq}$). In our notation, $F$ refers to logical fermions, $f$ refers to emergent ``physical" fermions, and $q$ refers to qubits. Fermions are defined on the vertices, and qubits are placed on the edges, with red and blue edges corresponding to Pauli $X$ and $Z$ operators acting on those qubits, respectively.}
    \label{fig:schematics}
\end{figure}

In this work, we introduce fermion-to-qubit stabilizer codes with arbitrarily large code distance for 2D and 3D fermionic systems. These codes preserve the locality of logical operators, and their stabilizers are local with constant weights, independent of both the code distance and the number of fermionic sites being simulated. The key idea is to perform a two-level concatenation, in which we first encode ``physical" fermions as quasiparticle excitations created on top of the long-range-entangled ground states of a qubit code with a small code distance $d_{fq}$~\cite{chen_exact_2018,chen_bosonization_2019, chen_error-correcting_2024}, and then encode logical fermions into these physical fermions through 2D fermionic color codes~\cite{bravyi2010majorana, vijay2015, litinski2018quantum, schuckert_fermion-qubit_2024} with a large code distance $d_{Ff}$. The overall distance $d_{Fq}$ scales as $\Theta(d_{Ff}d_{fq})$, where increasing $d_{Ff}$ as desired yields the arbitrarily large distance. To encode multiple fermions, we deform and embed multiple blocks of 2D fermionic color codes onto a square or cubic lattice. The lattice of logical fermions can then be used for quantum simulations of 2D and 3D systems (see Fig.~\ref{fig:schematics}).

Existing 3D locality-preserving fermion-to-qubit mappings have minimal or no error-correcting properties. Our work presents the first construction that combines high distance, constant stabilizer weight, and locality preservation in 3D. While prior approaches have achieved notable progress in extending the distances of fermion-to-qubit mappings, some key properties remain absent. Ref.~\cite{chen_error-correcting_2024} presents 2D locality-preserving mappings with a high encoding rate (the ratio of encoded logical fermionic modes to physical qubits) and a systematic method to numerically search for codes with increased distances. However, their stabilizer weights grow with distance, making experimental realizations challenging. An alternative method proposed in Ref.~\cite{iv2024low} numerically optimizes mappings with higher distances for hardware, but their code distance can only reach seven due to the numerical complexity. Locality-preserving codes with arbitrary code distances and low stabilizer weights have been constructed for simulating 2D fermionic systems by pulling apart twist defects that encode fermionic sites~\cite{landahl2023logicalfermionsfaulttolerantquantum,algaba_fermion-qubit_2025}. However, generalizing this construction to 3D fermionic systems remains challenging, and no known codes achieve these desirable properties in 3D. Our 2D code, based on a different construction, exhibits the same code properties and achieves the same scaling of the encoding rate as the codes in Refs.~\cite{landahl2023logicalfermionsfaulttolerantquantum,algaba_fermion-qubit_2025}, and, crucially, can be generalized to 3D while preserving these properties. In doing so, our 3D code directly addresses this gap and provides a robust, scalable framework for simulating 3D fermionic systems.

Ref.~\cite{algaba_fermion-qubit_2025} raises a broader concern that constructing high-distance fermion-to-qubit codes through concatenation typically results in high-weight stabilizers~\cite{derby2021compact,verstraete2005mapping,ball2005fermions, chiew2023discovering, miller2023bonsai, jiang2020optimal, Vlasov_2022, setia2019superfast}. Nevertheless, a notable counterexample to this observation is the concatenation of the Jordan-Wigner transformation with high-distance qubit surface codes. In this case, the code distance after concatenation is guaranteed by the distance of the logical qubits of the surface code, while the stabilizer weights remain low because they consist solely of the surface code stabilizers. However, the nonlocality problem persists for higher-dimensional fermionic systems in this approach. By contrast, our concatenation construction not only ensures low-weight stabilizers but also preserves locality. It provides another counterexample to the original observation.

The remainder of this article is organized as follows. In Sec.~\ref{sec:2}, we introduce our 2D fermion-to-qubit encoding protocol and analyze the properties of the resulting code, including the stabilizers, the logical operators, the distance, and the encoding rate. In Sec.~\ref{sec:3}, we present the 3D generalization of our construction, highlighting the modifications needed to embed logical fermions in three dimensions. Finally, in Sec.~\ref{sec:dis}, we discuss potential extensions of our codes, propose decoders for stochastic noise, and compare the encoding rate to that of previous work.

\section{2D encoding protocol}\label{sec:2}
Following the outline in Sec.~\ref{sec:overview} and Fig.~\ref{fig:schematics}, we first restrict the Pauli algebra to the subalgebra of emergent ``physical'' fermions on the toric code. To enforce this restriction, stabilizers are introduced to detect low-weight Pauli errors that do not correspond to fermionic excitations~\cite{chen_exact_2018,chen_bosonization_2019, chen_error-correcting_2024}. Next, to extend the overall code distance, we construct additional stabilizers to encode logical fermions into physical fermions through 2D fermionic color codes \cite{bravyi2010majorana, vijay2015, litinski2018quantum, schuckert_fermion-qubit_2024}. These stabilizers are designed to detect errors that can be mapped to fermionic excitations. In the end, we combine multiple blocks of 2D fermionic color codes onto the same lattice. 

Throughout this work, we employ Majorana operators to represent fermions. Specifically, the anticommutation relation for fermions is $\{c_i,c_j^\dagger\}=\delta_{ij}$ where $c_i$ and $c_j^\dagger$ are the fermionic lowering and raising operators at vertices $i$ and $j$, respectively. The Majorana operators $\gamma_i, \tilde{\gamma}_i$  are defined as $\gamma_j = c_j + c_j^\dagger$ and $\tilde\gamma_j = -i (c_j - c_j^\dagger)$ which satisfies $\{\gamma_i,\gamma_j\}=2\delta_{ij}=\{\tilde{\gamma}_i,\tilde{\gamma}_j\}$ and $\{\gamma_i,\tilde{\gamma}_j\}=0$. Given this relation, we use the words ``Majorana operators" and ``fermions" interchangeably in this paper.
\subsection{Small-distance mapping from physical fermions to qubits}\label{sec:chen2D}
The physical fermionic operators and spin operators can be related through local mappings (homomorphism) that preserve commutation and anticommutation relations. Throughout this section, we use one such mapping in 2D with $d_{fq} = 2$~\cite{chen_exact_2018, chen_error-correcting_2024}, illustrated below, as an example. We note, however, that our method and results naturally extend to the mappings with any $d_{fq}$ described in Refs.~\cite{chen_exact_2018, chen_error-correcting_2024}, with the caveat that a larger $d_{fq}$ comes at the expense of higher stabilizer weights. 

In this paper, qubits are placed on the edges of a square or cubic lattice, with red and blue edges corresponding to Pauli $X$ and $Z$ operators acting on those qubits, respectively. The fermions are defined on the vertices. The mapping of the operators is as follows~\cite{chen_exact_2018,chen_error-correcting_2024}: The hopping-up operator (also called a ``transfer operator"~\cite{algaba_fermion-qubit_2025}), in the direction from vertex $a$ to vertex $d$, is
\begin{equation}
\label{eq:hoppingup}
i\gamma_a\tilde{\gamma}_d \enspace \longleftrightarrow \enspace T_{ad}:=
\vcenter{\hbox{
\begin{tikzpicture}[
    scale=1.5,
    every node/.style={font=\small},
    thick,
    dot/.style={circle, fill=black, inner sep=1.5pt}
]

\node[dot] (a) at (1,1) {};
\node (b) at (2,1) {};
\node[dot] (d) at (1,2) {};

\node[below] at (a) {$a$};
\node[above] at (d) {$d$};

\draw[red]  (a) -- (d);
\draw[blue] (a) -- (b);

\end{tikzpicture}
}}
,
\end{equation}
while the hopping-right operator, in the direction from vertex $a$ to vertex $b$, is 
\begin{equation}
i\gamma_a\tilde{\gamma}_b\enspace\longleftrightarrow\enspace T_{ab}:=
\label{eq:hoppingright}
\vcenter{\hbox{
\begin{tikzpicture}[
    scale=1.5,
    every node/.style={font=\small},
    thick,
    dot/.style={circle, fill=black, inner sep=1.5pt}
]

\node[dot] (a) at (1,1) {};
\node[dot] (b) at (2,1) {};
\node (c) at (2,0) {}; 

\node[left] at (a) {$a$};
\node[right] at (b) {$b$};

\draw[red]  (a) -- (b);  
\draw[blue] (b) -- (c);  

\end{tikzpicture}
}}.
\end{equation}
The occupation operator (also called a ``vertex operator"~\cite{derby2021compact,algaba_fermion-qubit_2025}) at vertex $a$ is defined as

\begin{equation}
-i\gamma_a \tilde{\gamma}_a \enspace\longleftrightarrow\enspace W_a:=
\label{eq:wa_operator} 
\vcenter{\hbox{
\begin{tikzpicture}[
    scale=1.5,
    every node/.style={font=\small},
    thick,
    dot/.style={circle, fill=black, inner sep=1.5pt}
]

\draw[blue, thick] (-1,0) -- (1,0);   
\draw[blue, thick] (0,-1) -- (0,1);   

\node[dot] at (0,0) {};

\node[above right] at (0,0) {$a$};

\end{tikzpicture}
}}.
\end{equation}

The operators in Eqs.~(\ref{eq:hoppingup})--(\ref{eq:wa_operator}), together with their translated (shifted) versions on all vertices, are sufficient to generate an even fermionic algebra, in which every operator is a product of an even number of $\gamma$ and $\tilde{\gamma}$. The operators in the even fermionic algebra (preserving fermionic parity symmetry) are sufficient to describe all interactions relevant for quantum simulations. Accordingly, we choose the direction of the hopping operators in the generators to be upwards and to the right:
\begin{center}
\begin{tikzpicture}
[
    scale=1.5,
    every node/.style={font=\small},
    arrow/.style={-Latex, thick},
    halfline/.style={thick},
    dot/.style={circle, fill=black, inner sep=1.5pt}
]

\foreach \x in {0,1,2} {
    \foreach \y in {0,1,2} {
        \draw[thick] (\x,\y) rectangle ++(1,1);

        \draw[arrow] (\x+0.35,\y) -- (\x+0.65,\y);

        \draw[arrow] (\x+0.35,\y+1) -- (\x+0.65,\y+1);

        \draw[arrow] (\x,\y+0.35) -- (\x,\y+0.65);

        \draw[arrow] (\x+1,\y+0.35) -- (\x+1,\y+0.65);
    }
}

\node[below left]  at (1,1) {a};
\node[below right] at (2,1) {b};
\node[above right] at (2,2) {c};
\node[above left]  at (1,2) {d};

\foreach \x in {0,1,2,3} {
    \draw[halfline] (\x,0) -- ++(0,-0.3);  
    \draw[halfline] (\x,3) -- ++(0,0.3);   
}
\foreach \y in {0,1,2,3} {
    \draw[halfline] (0,\y) -- ++(-0.3,0);  
    \draw[halfline] (3,\y) -- ++(0.3,0);   
}

\foreach \x in {0,1,2,3} {
    \foreach \y in {0,1,2,3} {
        \node[dot] at (\x,\y) {};
    }
}
\end{tikzpicture}
\end{center}

From the generators above, we can also obtain the operator for hopping down as $i\gamma_d \tilde{\gamma}_a \propto\gamma_d(\tilde{\gamma}_d\tilde{\gamma}_d)(\gamma_a\gamma_a)\tilde{\gamma}_a \leftrightarrow W_d T_{ad}W_a$

\begin{equation}
i\gamma_d \tilde{\gamma}_a \enspace\longleftrightarrow \enspace T_{da}:=
\label{eq:hoppingdown}
\vcenter{\hbox{
\begin{tikzpicture}[
    scale=1.5,
    every node/.style={font=\small},
    thick,
    dot/.style={circle, fill=black, inner sep=1.5pt}
]

\draw[red, thick] (a) -- (d);

\draw[blue, thick] (1,0) -- (1,1);   
\draw[blue, thick] (1,2) -- (1,3);   

\draw[blue, thick] (0,1) -- (1,1);   
\draw[blue, thick] (0,2) -- (1,2);   
\draw[blue, thick] (1,2) -- (2,2);   

\node[dot] (a) at (1,1) {};
\node[dot] (d) at (1,2) {};

\node[right] at (a) {$a$};
\node[above right] at (d) {$d$};

\end{tikzpicture}
}},
\end{equation}

and for hopping left as

\begin{equation}
i\gamma_b \tilde{\gamma}_a \enspace\longleftrightarrow\enspace T_{ba}:=
\label{eq:hoppingleft}
\vcenter{\hbox{
\begin{tikzpicture}[
    scale=1.5,
    every node/.style={font=\small},
    thick,
    dot/.style={circle, fill=black, inner sep=1.5pt}
]

\draw[red, thick] (a) -- (b);

\draw[blue, thick] (1,0) -- (1,2);   
\draw[blue, thick] (2,1) -- (2,2);   

\draw[blue, thick] (0,1) -- (1,1);   
\draw[blue, thick] (2,1) -- (3,1);   

\node[dot] (a) at (1,1) {};
\node[dot] (b) at (2,1) {};

\node[below left] at (a) {$a$};
\node[below] at (b) {$b$};

\end{tikzpicture}
}}.
\end{equation}

It can be verified that the physical Majorana representation and the spin representation satisfy the same commutation and anticommutation rules: Two hopping operators anticommute if they share the same starting or ending vertex; otherwise, they commute. An occupation operator at a vertex only anticommutes with the hopping operators that start from that vertex or end at that vertex. The signs of the mapping do not matter since they can always be resolved by classically redefining the signs of the syndromes. Therefore, in this work, we ignore those signs, the ordering of Majorana operators, and the ordering of $X$ and $Z$ operators acting on the same edge in all operators.

An identity between Majorana operators leads to the constraint on spins %
\begin{equation}\label{eq:gaugeconstraint}
    \begin{aligned}
    1=&\left( i\gamma_d \tilde{\gamma}_c \right)
\left( i\gamma_b \tilde{\gamma}_c \right)
\left( i\gamma_a \tilde{\gamma}_d \right)
\left( i\gamma_a \tilde{\gamma}_b \right)
\left( -i\gamma_b \tilde{\gamma}_b \right)
\left( -i\gamma_d \tilde{\gamma}_d \right) \\
\leftrightarrow&T_{dc}T_{bc}T_{ad}T_{ab}W_bW_d,
    \end{aligned}
\end{equation}
That is, we have the stabilizer $G_d$ for this vertex $d$ defined as
\begin{equation}
G_d:= T_{dc}T_{bc}T_{ad}T_{ab}W_bW_d =
\vcenter{\hbox{
\begin{tikzpicture}[
    scale=1.5,
    every node/.style={font=\small},
    thick,
    redline/.style={red, thick},
    blueline/.style={blue, thick},
    dot/.style={circle, fill=black, inner sep=1.5pt}
]

\draw[blueline] (-1,0) -- (1,0);  
\draw[blueline] (0,-1) -- (0,1);  

\draw[redline] (1,-0.02) -- (1,-1);   
\draw[redline] (1,-1) -- (0.02,-1);   
\draw[redline] (0.02,-0.02) -- (0.020,-1.0);   
\draw[redline] (1,-0.02) -- (0.02,-0.02);   

\node[dot] at (0,0) {};
\node[above right] at (0,0) {$d$};

\end{tikzpicture}
}}=1.
\end{equation}

We can similarly define a stabilizer $G_v$ at each vertex $v$ by translating Eq.~(\ref{eq:gaugeconstraint}) from $d$ to $v$. For simplicity of notation, we use $G$ to refer to $G_v$ at all vertices $v$, and similarly for $T$ and $W$. Stabilizers $G$ form a $d_{fq}=2$ stabilizer code with operators $T$ and $W$~\cite{chen_exact_2018, chen_error-correcting_2024}.

The mapping from fermionic algebra to Pauli algebra corresponds to a mapping from fermionic Hilbert space to qubit Hilbert space. The fermionic Hilbert space is a subspace of the qubit Hilbert space and is stabilized by $G$. 

The stabilizers $G$ allow for the detection of Pauli errors that do not belong to the fermionic algebra, provided their weight is less than the distance $d_{Fq}$. It can be shown that if a Pauli operator commutes with $G$, then it either can be mapped to a product of Majorana operators $\gamma$ and $\tilde{\gamma}$, or corresponds to a (homologically) nontrivial loop of Pauli $Z$ operators, up to multiplication by stabilizers. Such a nontrivial loop is not the boundary of a surface and thus is not a product of plaquette stabilizers. This result can be physically interpreted in terms of anyon excitations in the 2D toric code, as $G$ is a product of $Z$-type star and $X$-type plaquette stabilizers of the 2D toric code. The ground states of the toric code, which are stabilized by these operators, are also stabilized by $G$. Using the convention that a Pauli $Z$ operator generates a pair of $m$-anyons on its neighboring plaquettes and a Pauli $X$ operator generates a pair of $e$-anyons on its neighboring vertices, we can see that a state is stabilized by all $G$ only when there are no anyons, or when the anyons appear in pairs with an $m$-anyon located to the lower right of an $e$-anyon. This type of $e$-$m$ anyon pair is fermionic due to its exchange statistics~\cite{chen_exact_2018}. A nontrivial loop of Pauli $Z$ operators, up to multiplication by stabilizers, does not create anyon excitations, but such a nontrivial loop can correspond to a logical error during the simulation of fermionic systems if it commutes with additional stabilizers introduced later. See Appendix~\ref{apx:z} for further discussion.

\subsection{Increasing distance and number of fermionic sites}
After obtaining the fermionic algebra, we extend the code distance $d_{Fq}$ by forming logical fermions $\gamma^L$ and $\tilde{\gamma}^L$ based on fermionic color codes~\cite{bravyi2010majorana, vijay2015, litinski2018quantum, schuckert_fermion-qubit_2024} built from emergent physical fermions $\gamma$ and $\tilde{\gamma}$. Each logical fermionic mode is encoded in a color code block and formed from physical fermionic modes positioned on the vertices of that code block. Unlike the conventional approach in qubit codes or fermionic codes with non-emergent fermions (see Fig.~3(c) in~\cite{schuckert_fermion-qubit_2024}), where each code block is placed on a separate lattice, we embed multiple code blocks into a shared underlying lattice to accommodate constraints imposed by emergent fermions. This shared-lattice embedding is crucial for preserving the anticommutation relations between logical fermions. Preserving the anticommutation relations requires that each logical fermion be formed from an odd number of physical fermions. However, fermion-parity symmetry forbids the presence of an odd number of emergent physical fermions on a lattice. Embedding multiple code blocks into a single lattice resolves this conflict, as the total number of emergent physical fermions remains even. Another way to describe this constraint is that if different code blocks were placed on separate lattices, each logical fermion would be supported solely by Pauli operators acting on its respective lattice. In such a case, the anticommutation relations between logical fermions could not be preserved under the mapping because Pauli operators supported on distinct qubit lattices necessarily commute with each other. Embedding all code blocks into a shared lattice avoids this issue without manually imposing the anticommutation relations on the definitions of logical operators.

In our construction, we deform 2D honeycomb color codes and embed multiple color code blocks into the same square lattice, as shown in Fig.~\ref{fig:code-blocks}. The deformation arises because the $d_{fq}=2$ code used in the concatenation is defined on square lattices. Although the architecture involves deformations, the arrangement of color code blocks makes our 2D code suitable for simulating fermionic systems on square lattices. Fermions on other lattices can be simulated as well by modifying the placement of the color code blocks in our code, as discussed in Sec.~\ref{sec:other}.

\begin{figure}[htbp]
    \centering
\includegraphics[width=1\linewidth]{embedding.pdf}
    \caption{(a) Embedding of a color code block into a square lattice. The numbering of corresponding vertices on the left side and on the right side demonstrates this deformation. The plaquette-type stabilizers are deformed in one of two ways, depending on the location of the color code block after embedding. (b) Multiple fermionic color code blocks are placed on a 2D lattice. Each code block can encode logical Majorana operators $\gamma^{(x,y)L}$ or $\tilde{\gamma}^{(x,y)L}$, which are defined in Eqs.~(\ref{eq:logcol1}) and (\ref{eq:logcol2}). Here, ${(x,y)}$ labels their locations, going from $(1,1)$ to the number of fermionic sites in each direction, and the superscript $L$ stands for logical operators. The code blocks are placed in staggered vertical columns labeled by $y$, with different deformations applied to blocks in odd and even columns to optimize the encoding rate. The numbering of vertices in (b) for all code blocks (both odd and even) corresponds to the numbering of vertices in the code block on the left side of (a) before embedding. In addition to the plaquette-type stabilizers, we also introduce padding stabilizers, defined as occupation operators $W_p = -i\gamma_p\tilde{\gamma}_p$ at the vertices, labeled $p$, in the space between code blocks. While this figure illustrates color code blocks of distance 5 as an example, the same principle can be applied to deform and embed color codes with larger distances.}
    \label{fig:code-blocks}
\end{figure}

\begin{figure*}[htbp]
  \centering
  \includegraphics[width=\textwidth]{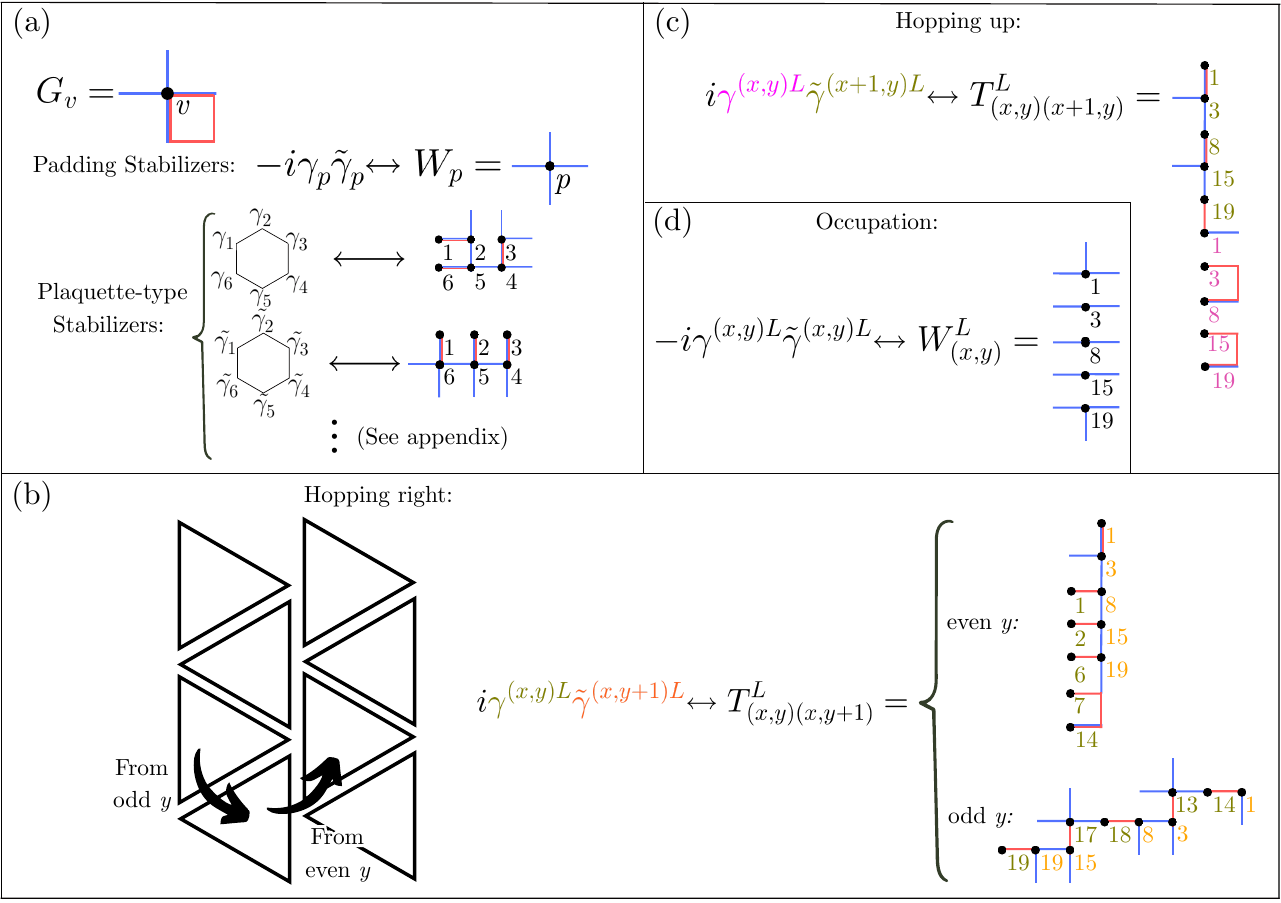}
  \caption{Mapping of the generators of stabilizers and logical operators in their minimum-weight representation, which may not be unique, for our 2D fermion-to-qubit code. (a) Stabilizer generators of the code. We have: (i) $G_v$ at all vertices $v$, used for the mapping from qubits to physical fermions; (ii) padding stabilizers located in the regions between color code blocks; and (iii) plaquette-type stabilizers on the color code mapped based on their shapes and locations (as shown in Fig.~\ref{fig:code-blocks}). The mapping of a full list of plaquette-type stabilizers is shown in Appendix~\ref{apx:stabilizers}. The numbering of vertices in (a) does not correspond to the numbers in Fig.~\ref{fig:code-blocks}. (b) Logical hopping-right operators hopping from the color code block at location $(x,y)$ to the color code block at $(x, y+1)$. The operator has two versions, depending on whether the starting point $y$ is even or odd. The color labels of vertices in $T^L$ match the color labels of $\gamma^L$ or $\tilde{\gamma}^L$. The numbering of the vertices in the logical operators corresponds to that in Fig.~\ref{fig:code-blocks}. (c) The logical hopping-up operator goes from a color code block at location $(x,y)$ to the one at location $(x+1,y)$, which is the one above it with the same deformation. (d) Logical occupation operator associated with the color code block at position $(x, y)$. The minimum-weight representations are verified by ensuring that applying stabilizers $G$ cannot further reduce the operator weights.}
  \label{fig:2dsummary}
\end{figure*}

\subsection{Mapping of stabilizers and logical operators}
To detect fermionic errors $E$, defined as errors that can be mapped to the fermionic algebra, we introduce additional stabilizers drawn in Fig.~\ref{fig:2dsummary} based on 2D color codes. These fermionic errors correspond to the residual errors that cannot be corrected by applying $G$. The support of $E$ within a color code block can be detected by fermionic color code stabilizers, which are the products of $\gamma$ (or $\tilde{\gamma}$) around each color code plaquette, as colored in Fig.~\ref{fig:code-blocks}(a). The ordering of Majorana operators does not affect their (anti)commutation relations with other operators and is therefore neglected. These plaquette-type stabilizers can be mapped to Pauli operators using the hopping and occupation operators of the physical fermions in Eqs.~(\ref{eq:hoppingup})--(\ref{eq:wa_operator}), and their weight can potentially be reduced using the stabilizers $G_v$. For example, one of the stabilizers would be mapped as follows:
\begin{equation}\label{eq:stabalizerexample}
    \begin{aligned}
    &\tilde{\gamma_{1}}\tilde{\gamma_{2}}\tilde{\gamma_{3}}\tilde{\gamma_{4}}\tilde{\gamma_{5}}\tilde{\gamma_{6}}\\
    = & -(-i\gamma_{1}\tilde{\gamma_{1}})(i\gamma_{1}\tilde{\gamma_{2}})(-i\gamma_{4}\tilde{\gamma_{4}})(i\gamma_{4}\tilde{\gamma_{3}})(-i\gamma_{6}\tilde{\gamma_{6}})(i\gamma_{6}\tilde{\gamma_{5}})\\
\leftrightarrow& W_1T_{12}W_4T_{43}W_6T_{65}.
    \end{aligned}    
\end{equation}

Pictorially, the mapping is
\begin{equation}\label{eq:stabalizerexample2}
\begin{tikzpicture}[baseline={(current bounding box.center)}, scale=1.5, every node/.style={font=\small}]
    \def\r{0.5}
    \coordinate (A) at (150:\r);
    \coordinate (B) at (90:\r);
    \coordinate (C) at (30:\r);
    \coordinate (D) at (-30:\r);
    \coordinate (E) at (-90:\r);
    \coordinate (F) at (-150:\r);
    \draw (A) -- (B) -- (C) -- (D) -- (E) -- (F) -- cycle;
    \node[above left]  at (A) {$\tilde\gamma_1$};
    \node[above]       at (B) {$\tilde\gamma_2$};
    \node[above right] at (C) {$\tilde\gamma_3$};
    \node[below right] at (D) {$\tilde\gamma_4$};
    \node[below]       at (E) {$\tilde\gamma_5$};
    \node[below left]  at (F) {$\tilde\gamma_6$};
\end{tikzpicture}
\leftrightarrow
\begin{tikzpicture}[baseline={(current bounding box.center)}, scale=1.2, every node/.style={font=\small}, 
    redline/.style={red, thick},
    blueline/.style={blue, thick},
    dot/.style={circle, fill=black, inner sep=1pt}]
    \coordinate (1) at (0,1);
    \coordinate (2) at (0.5,1);
    \coordinate (3) at (1,1);
    \coordinate (4) at (1,0.5);
    \coordinate (5) at (0.5,0.5);
    \coordinate (6) at (0,0.5);
    \draw[blueline] (0, 1.03) -- (0.5, 1.03);
    \draw[blueline] (0, 0.53) -- (0.5, 0.53);
    \draw[blueline] (1.03, 1) -- (1.03, 0.5);
    \draw[blueline] (0,1.5) -- (0,1);      
    \draw[blueline] (0,0) -- (0,0.5);
    \draw[blueline] (0.5,1) -- (0.5,0);    
    \draw[blueline] (1,0.5) -- (1,0);
    \draw[blueline] (-0.5,1) -- (0.5,1);     
    \draw[blueline] (-0.5,0.5) -- (0.75,0.5); 
    \draw[blueline] (0.75,0.5) -- (1,0.5);    
    \draw[redline] (1) -- (2);
    \draw[redline] (6) -- (5);
    \draw[redline] (3) -- (4);
    \node[dot] at (1) {};
\node at ([xshift=3pt, yshift=-4pt]1) {$1$};

\node[dot] at (2) {};
\node at ([xshift=3pt, yshift=-4pt]2) {$2$};

\node[dot] at (3) {};
\node at ([xshift=3pt, yshift=-4pt]3) {$3$};

\node[dot] at (4) {};
\node at ([xshift=3pt, yshift=-4pt]4) {$4$};

\node[dot] at (5) {};
\node at ([xshift=3pt, yshift=-4pt]5) {$5$};

\node[dot] at (6) {};
\node at ([xshift=3pt, yshift=-4pt]6) {$6$};

\end{tikzpicture}
 = 
\begin{tikzpicture}[baseline=0.9cm, scale=1.2, every node/.style={font=\small}, 
    redline/.style={red, thick},
    blueline/.style={blue, thick},
    dot/.style={circle, fill=black, inner sep=1pt}]

    \coordinate (1) at (0,1);
    \coordinate (2) at (0.5,1);
    \coordinate (3) at (1,1);
    \coordinate (4) at (1,0.5);
    \coordinate (5) at (0.5,0.5);
    \coordinate (6) at (0,0.5);

    \draw[blueline] (1.03, 1) -- (1.03, 0.5);   
    \draw[blueline] (0.53, 1) -- (0.53, 0.5);  
    \draw[blueline] (0.03, 1) -- (0.03, 0.5);  
    \draw[blueline] (0,0) -- (0,0.5);
    \draw[blueline] (0.5,1) -- (0.5,0);    
    \draw[blueline] (1,0.5) -- (1,0);
    \draw[blueline] (-0.5,0.5) -- (0.75,0.5); 
    \draw[blueline] (0.75,0.5) -- (1,0.5);    

    \draw[redline] (1) -- (6);
    \draw[redline] (2) -- (5);
    \draw[redline] (3) -- (4);

    \node[dot] at (1) {};
    \node at ([xshift=3pt, yshift=-4pt]1) {$1$};
    \node[dot] at (2) {};
    \node at ([xshift=3pt, yshift=-4pt]2) {$2$};
    \node[dot] at (3) {};
    \node at ([xshift=3pt, yshift=-4pt]3) {$3$};
    \node[dot] at (4) {};
    \node at ([xshift=3pt, yshift=-4pt]4) {$4$};
    \node[dot] at (5) {};
    \node at ([xshift=3pt, yshift=-4pt]5) {$5$};
    \node[dot] at (6) {};
    \node at ([xshift=3pt, yshift=-4pt]6) {$6$};
\end{tikzpicture}
,
\end{equation}
where in the last step, we applied the stabilizer $G_1$ to reduce the weight from $11$ to $9$.

We can see that each plaquette-type stabilizer is a product of operators $T$ and $W$, all of which commute with the previously defined stabilizers $G$. Hence, each plaquette-type stabilizer also commutes with $G$, satisfying the requirement for being a stabilizer.

In addition, we detect Majorana operators residing on the vertices between different code blocks (the “padding”) by placing an occupation operator $W_p = -i\gamma_p\tilde{\gamma}_p$ on each padding vertex $p$ as a stabilizer. $W_p$ commutes with all previously defined stabilizers since none of the hopping or occupation operators in the plaquette-type stabilizers involve Majorana operators at vertex $p$. By including $W_p$ as stabilizers, we eliminate $\gamma_p$ or $\tilde{\gamma}_p$ from the code space. Our current staggered configuration of color code blocks in Fig.~\ref{fig:code-blocks} results in $(d_{Ff}-1)/4$ padding vertices per color code block, which is relatively efficient in terms of encoding rate. The encoding rate can be slightly improved by deforming each color code block differently to eliminate all gaps between them. However, this results in logical operators and stabilizers having many different configurations depending on the deformation, making them harder to track.

The logical operators we have for fermionic simulation are the logical hopping operators for the logical Majorana operators from a color code block $a$ to another code block $b$, $T_{ab}^L = i \gamma^{aL}\tilde{\gamma}^{bL}$, and the logical occupation operator $W_{a}^L = i \gamma^{aL}\tilde{\gamma}^{aL}$ at a code block $a$. Taking color code blocks of distance $d_{Ff}=5$ as an example, the logical Majorana operators on a block 
$a$ are defined based on the numbering of the vertices in Fig.~\ref{fig:code-blocks}(b) as 
\begin{align}
\begin{split}
    \gamma^{aL} &= \gamma_1^a \gamma_3^a \gamma_8^a \gamma_{15}^a \gamma_{19}^a, \\
&= \gamma_{14}^a \gamma_{13}^a \gamma_{18}^a \gamma_{17}^a \gamma_{19}^a,\\
&= \gamma_{1}^{a} \gamma_{2}^{a} \gamma_{6}^{a} \gamma_{7}^{a} \gamma_{14}^{a},\label{eq:logcol1}
\end{split}\\
\begin{split}
    \tilde{\gamma}^{aL} &= \tilde{\gamma}_{1}^a \tilde{\gamma}_{3}^a \tilde{\gamma}_8^a \tilde{\gamma}_{15}^a \tilde{\gamma}_{19}^a, \\
&= \tilde{\gamma}_{14}^a \tilde{\gamma}_{13}^a \tilde{\gamma}_{18}^a \tilde{\gamma}_{17}^a \tilde{\gamma}_{19}^a,\\
&=\tilde{\gamma}_{1}^a \tilde{\gamma}_{2}^a \tilde{\gamma}_{6}^a \tilde{\gamma}_{7}^a \tilde{\gamma}_{14}^a,
\end{split}\label{eq:logcol2}
\end{align}
up to stabilizers and signs. The mapping of the logical operators in 2D from logical Majorana operators to qubits is explicitly shown in Fig.~\ref{fig:2dsummary}. We observe that local fermionic operators are mapped to local spin operators, and this property holds regardless of the number of fermionic sites.

\subsection{Scaling behavior of distance and encoding rate}\label{sec:dproof}
Our key contribution is that while preserving locality, the code distance can be extended to infinity without increasing the stabilizer weight. In our example for $d_{fq}=2$, the operator weights are $|T_{ab}^L| \geq 
\frac{5}{2}d_{Ff}-\frac{1}{2}$ for hopping right or up, and $|W_{a}^L|=2d_{Ff}+2$ for occupation operators. In general, the weights of the logical operators, and therefore the code distance, scale as \begin{equation}
    d_{Fq} = \Theta(d_{Ff} d_{fq}).
\end{equation}
Using the construction we proposed, the overall distance can be increased by increasing either $d_{fq}$, as proposed in Ref.~\cite{chen_error-correcting_2024}, or $d_{Ff}$. The approach of increasing $d_{fq}$ comes with the disadvantage of increasing stabilizer weights, but increasing the color code distance $d_{Ff}$ improves the overall code distance while keeping the stabilizer weights small—on the order of $d_{fq}$. This property is crucial for practical experimental realizations.

We prove that cancellations between the logical operators and the stabilizers do \textit{not} reduce the scaling of the distance and thus $d_{Fq} = \Theta(d_{Ff})$ for a fixed $d_{fq}$. This is established as follows: An increase in the fermionic color code distance $d_{Ff}$ implies that all logical operators must be composed of more physical fermions, since by definition they act on at least $d_{Ff}$ physical fermions. In the physical-fermion-to-qubit mappings we used~\cite{chen_exact_2018, chen_bosonization_2019, chen_error-correcting_2024} with a fixed $d_{fq}$, generating more physical fermions requires a greater number of Pauli operators. This follows from the fact that for a fermion to occupy a vertex, Pauli operators must act on at least one of the edges connected to that vertex. Consequently, the number of Pauli operators grows at least linearly with the number of physical fermions encoded, leading to a lower bound of $d_{Fq} = \Omega(d_{Ff})$. Moreover, the code distance is bounded above by $\mathcal{O}(d_{Ff})$ since a linear expression relating the weight of the logical occupation operator $|W_a^L|$ to $d_{Ff}$ was shown above for $d_{fq} = 2$, and similar expressions can be derived for other $d_{fq}$. Together, these bounds establish $d_{Fq} = \Theta(d_{Ff})$ for a fixed $d_{fq}$ in our code.

The encoding rate of our 2D code, defined as the number of logical fermionic sites per qubit, scales as $\Theta(1/d_{Fq}^2)$, due to the scaling behavior of 2D color codes. The encoding rate is independent of $d_{fq}$, as the encoding rate for physical fermions remains fixed at $1/2$ for any $d_{fq}$, given that the physical fermions are defined on the vertices of a square lattice and the qubits are associated with its edges. In Sec.~\ref{sec:enc}, we discuss how our encoding rate compares with that of previous work.

\section{3D encoding protocol}\label{sec:3}
\begin{figure}[htbp]
    \centering
    \includegraphics[width=\linewidth]{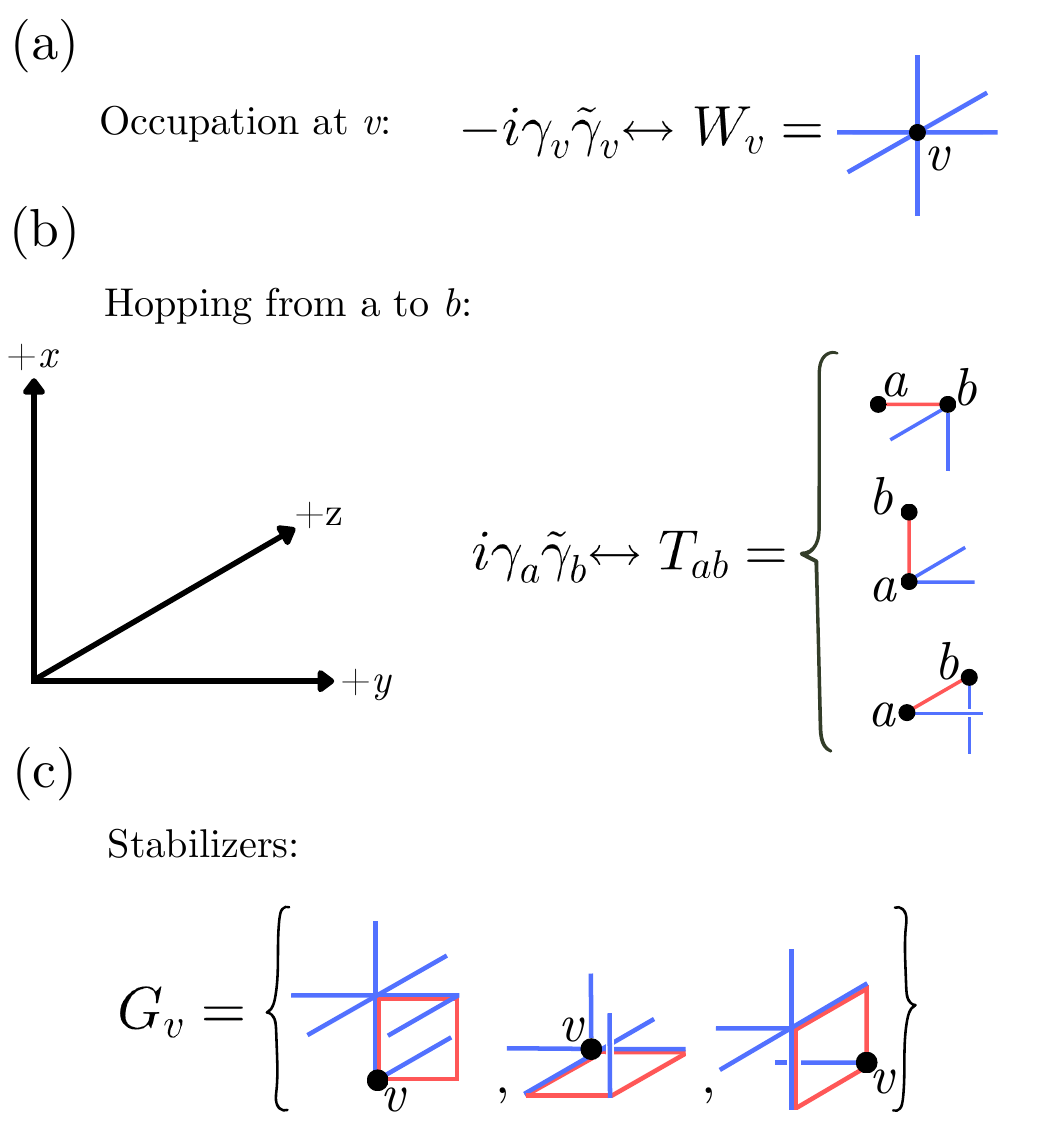}
    \caption{A mapping between 3D Pauli operators and the corresponding logical operators and stabilizers in the Majorana basis: (a) a physical occupation operator $W_v$ defined on each vertex $v$; (b) physical hopping operators $T_{ab}$ hopping in the positive directions, where the $+z$ direction is into the page; (c) three stabilizers corresponding to the constraints of Majorana algebra, similar to Eq.~(\ref{eq:gaugeconstraint}), on three 2D planes. Majorana operators are defined on vertices, while qubits are defined on edges. Adapted from Ref.~\cite{chen_bosonization_2019}.}\label{fig:3dmapping}
\end{figure}

Our 3D encoding protocol follows the same framework as our 2D protocol, as illustrated in Fig.~\ref{fig:schematics}, but we make two changes:
\begin{enumerate}
    \item On the level of creating emergent physical fermions from qubits, we use a 3D mapping with distance $d_{fq}=3$ in Ref.~\cite{chen_bosonization_2019}. The stabilizers, hopping and occupation operators on this level are shown in Fig.~\ref{fig:3dmapping}.
    \item To simulate a 3D fermionic system on a cubic lattice, we stack multiple layers—each consisting of the 2D staggered arrangement of fermionic color codes shown in Fig.~\ref{fig:code-blocks}—along the $z$-direction. The distance of the formed ``stacked" code, $d_{Ff}$, is identical to the distance of one 2D color code block.
\end{enumerate}

The stabilizers and logical operators of our 3D encoding protocol are summarized in Fig.~\ref{fig:3dsummary}.

\subsection{Detection of Pauli errors outside fermionic algebra in 3D}

In 2D, the physical interpretation of anyon excitations leads to the conclusion that any Pauli error not in the fermionic algebra either anticommutes with $G$ or, if it commutes with $G$, must be in the form of a nontrivial loop of Pauli $Z$ operators (up to stabilizers). As this interpretation does not readily generalize to 3D, we construct an alternative proof as follows.

Since Pauli $X$ operators can be generated from the hopping operators $T$ and Pauli $Z$ operators, $\langle T, Z\rangle=\langle I, X, Z, Y\rangle$ generates the space of all Pauli operators on the lattice. Moreover, since $[T,G]=0$, the commutation and anticommutation relation between a Pauli operator $O$ and $G$ only depends on the Pauli $Z$ component of $O$. The $X$-component of $G$ is identical to the plaquette stabilizers of the 3D toric code~\cite{kitaev_fault-tolerant_2003}. Therefore, a product of Pauli $Z$ operators at various locations commutes with $G$ if and only if those operators form a closed membrane under periodic boundary conditions. There are two possibilities: (i) the membrane of Pauli $Z$ operators is equal to a product of occupation operators $W$, and thus belongs to the fermionic algebra; (ii) the membrane is topologically nontrivial, crossing the entire lattice (which includes multiple code blocks), and cannot be expressed as a product of $T$ and $W$. However, such a nontrivial membrane has high Pauli weight, and the probability of its occurrence is effectively negligible. More discussion on this can be found in Appendix~\ref{apx:z}. We thus conclude that, excluding the contribution from nontrivial $Z$ membranes, any Pauli operator that commutes with $G$ must be composed of $T$ and $W$, or equivalently, can be mapped to $\gamma$ and $\tilde{\gamma}$, and therefore belongs to the fermionic algebra.

\subsection{3D code distance and encoding rate}
The 3D code is locality-preserving and the stabilizer weights remain constant as the distance increases. Similar to the 2D case, the 3D code allows for arbitrarily large distance $d_{Fq}=\Theta(d_{Ff} d_{fq})$, as the proof of distance scaling Sec.~\ref{sec:dproof} remains valid in 3D. The encoding rate scaling $\Theta(1/d_{Fq}^2)$ is identical to that of the 2D construction, since each logical fermion is still built from a 2D fermionic color code with $\Theta(d_{Ff}^2)$ physical fermions, and the 2D layers can be stacked in 3D without requiring additional spacing between them to preserve the code distance.

\begin{figure*}[htbp]
    \centering
    \includegraphics[width=\textwidth]{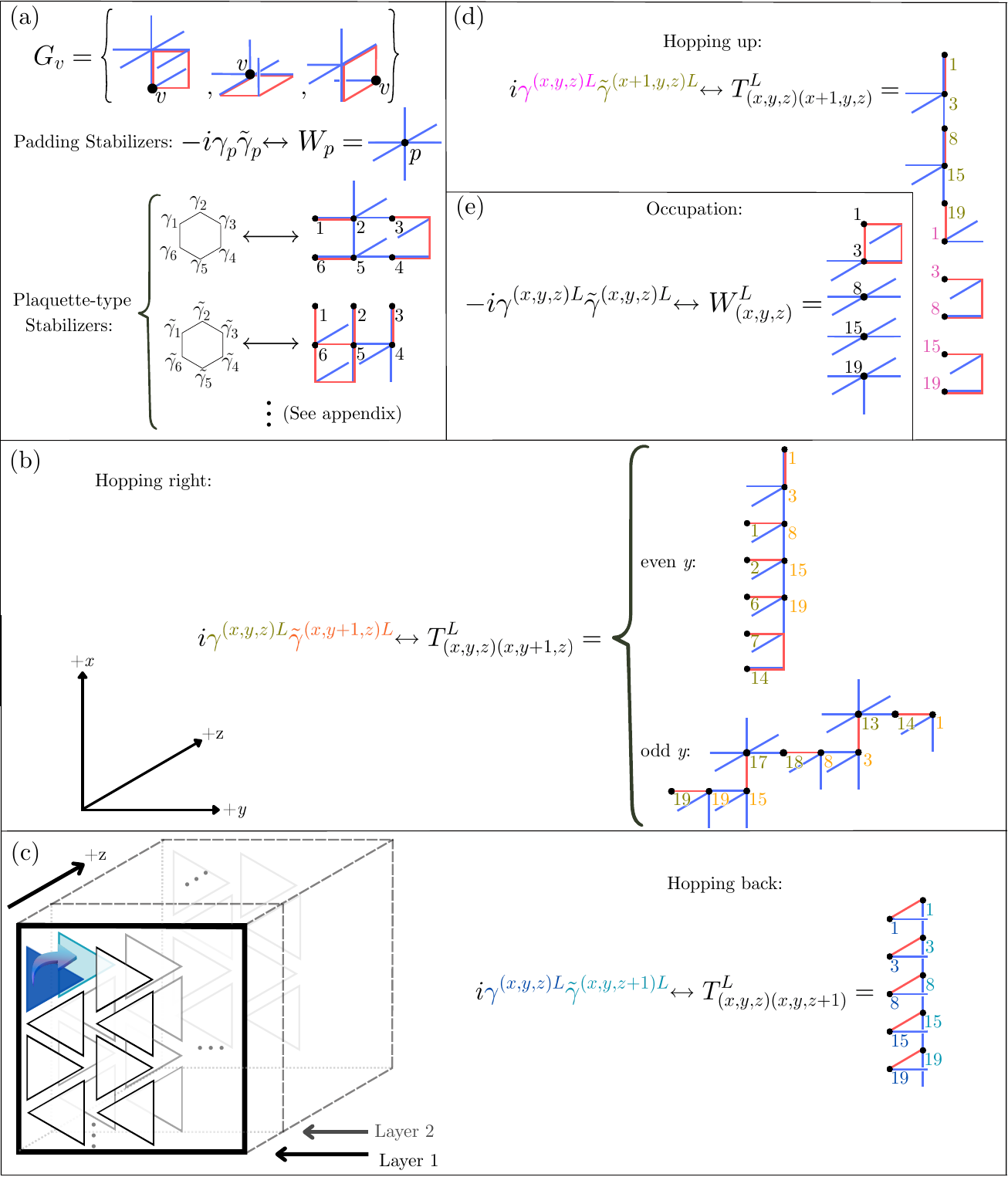}
    \caption{Summary of the generators of stabilizers and logical operators for our 3D fermion-to-qubit code. (a) Stabilizer generators of the code. We have: (i) Three stabilizers $G_v$ at each vertex $v$; (ii) padding stabilizers between the color code blocks on each layer; and (iii) plaquette-type stabilizers on the color code mapped based on their shapes and locations (as shown in Fig.~\ref{fig:code-blocks}). All plaquette-type stabilizers for 3D code are shown in Appendix~\ref{apx:stabilizers}. (b)(c)(d) Logical hopping operators hopping from the color code block at location $(x,y)$. The logical hopping back operator hops from the color code block at $(x, y, z)$ to the block at $(x, y, z+1)$, which is positioned directly behind it in the next layer. (e) Logical occupation operator associated with the color code block at position $(x, y)$.}
    \label{fig:3dsummary}
\end{figure*}

\section{Discussion and outlook}\label{sec:dis}

In conclusion, we constructed high-distance fermion-to-qubit codes in 2D and 3D by concatenating 2D fermionic color codes with 2D or 3D small-distance fermion-to-qubit codes. In both cases, the code distance scales linearly with the color code distance. The stabilizers remain local with constant weight regardless of the code distance. The logical operators are locality-preserving when the code distance is fixed, and the stabilizers and logical operators for the 2D and 3D constructions are summarized in Figs.~\ref{fig:2dsummary} and~\ref{fig:3dsummary}, respectively. In addition, the encoding rates for both codes obey inverse-square scaling in the code distance.

The implications of our work and areas for future research include:

\subsection{Adaptation to alternative fermionic systems and qubit hardware geometries}\label{sec:other}

Our code construction is based on simulating fermionic systems on square or cubic lattices using qubits placed on the same type of lattice. However, the codes can be tailored to alternative fermionic systems or qubit geometries. For instance, a fermionic system on a square lattice can be simulated using qubits on a 2D honeycomb lattice by replacing the physical-fermion-to-qubit encoding in the concatenation protocol with the small-distance encoding proposed in Ref.~\cite{chen_exact_2018}. Conversely, a fermionic system on a honeycomb lattice can be simulated using a square qubit lattice by adapting the embedding step in Fig.~\ref{fig:code-blocks} to arrange the logical fermion code blocks into a honeycomb geometry. This flexibility is particularly useful for adapting the code to experimental implementations where the available qubit hardware imposes specific geometric constraints.

In addition, our codes assume spinless fermions. Nevertheless, spinful fermions (fermions with spin) can also be simulated using our codes by interpreting multiple neighboring logical fermion code blocks as components of a single site and assigning a distinct spin label to each logical fermion within that site. For example, to simulate spin-half fermions, we can define logical fermions at odd and even columns as spin-up and spin-down, respectively, and construct the new logical nearest-site hopping operators from the original logical hopping and occupation operators.

\subsection{Decoding}\label{sec:decoding}

The development of effective decoders for fermion-to-qubit error-correcting codes remains an open problem. With the lack of effective decoding methods in quantum error correction, quantum error mitigation techniques, such as post-selection and extrapolation, have been explored as an alternative to suppress logical errors~\cite{bonet2018,chien2023simulating,Nigmatullin_2025,papič2025neartermfermionicsimulationsubspace}. However, the high sampling overhead associated with quantum error mitigation hinders the performance of such methods beyond simulating fermionic systems of small size or executing on qubit systems with very low physical error rates~\cite{dyrenkova2025scalablesimulationfermionicencoding}.

Our high-distance error-correcting codes can serve as good quantum memories for storing the results of fermionic quantum simulations without relying on quantum error mitigation, making these codes suitable for the simulation of larger systems. The rationale is as follows.

Noise in a qubit system is typically modeled using either an adversarial error model, which assumes that all errors of a fixed weight may occur in their worst-case configuration, or a stochastic error model, in which each qubit is independently subject to a Pauli $X$ or $Z$ error, each occurring with a small probability. In an adversarial error model, the large distance of our code $d_{Fq}$ guarantees that any error of weight less than or equal to $\lfloor \frac{d_{Fq}-1}{2} \rfloor$ can be corrected~\cite{nielsen_quantum_2010}. However, in a stochastic error model, the total error weight typically scales linearly with the number of qubits and is likely to exceed the code distance. A decoder with a threshold is required to correct errors in large systems, yet discussions of decoders have been lacking in previous literature on high-distance fermion-to-qubit codes.

Since our construction is based on code concatenation, we propose to decode by concatenating the decoders of the component codes. The small-distance physical-fermion-to-qubit code can be decoded using a lookup table for single-qubit errors (e.g., Fig.~(2) of Ref.~\cite{chen_error-correcting_2024}). Errors on multiple qubits in close proximity may be mapped to correlated physical fermionic errors, but such correlations decay exponentially with spatial separation. We correct these physical fermionic errors by adapting a decoder for 2D qubit color codes to 2D fermionic color codes. By choosing a suitable qubit color code decoder, we expect the existence of a threshold, below which the logical fermionic error rate on each logical fermion code block is exponentially suppressed with the overall code distance. Furthermore, we argue that increasing the number of logical fermionic sites $N_F$ does not destroy the threshold if $d_{Ff} = \Omega(\log N_F)$. The details of the decoder construction and the threshold argument are provided in Appendix~\ref{sec:decoder_details}. A rigorous proof and numerical simulations for the thresholds are left for future work.

In the argument for the existence of a threshold, we assumed perfect measurements. However, measurement errors have been a major obstacle in the decoding process of quantum error correction~\cite{bombin2015single}. Single-shot error correction is a highly desired property, as repeated measurements are time-consuming and could lead to more physical errors during the process. It was previously proven in Ref.~\cite{quintavalle2021single} that confinement is a sufficient condition for the existence of a single-shot decoder. Confinement means that a larger error will anticommute with more stabilizers. In the 3D toric code, only the type of error corresponding to loop-like excitations has confinement, since a string of the other type of error will only violate vertex stabilizers at the end of the string, regardless of the length of the string~\cite{dennis2002topological}. In contrast, when we map qubit errors to fermionic errors in our 3D code, both Pauli $X$ and $Z$ errors will be confined~\cite{zhou2025finite}, and as a result, the measurement error outside of the fermionic algebra can be corrected via single-shot decoding. However, we note that within the fermionic algebra, multiple rounds of measurements are still required to correct measurement errors of the plaquette-type stabilizers, since 2D color codes do not have confinement.

\subsection{Encoding rate comparison}\label{sec:enc}
In 2D, our encoding rate scales as $\Theta(1/d_{Fq}^2)$. This asymptotic scaling is identical to that of the 2D encodings in Refs.~\cite{algaba_fermion-qubit_2025} and \cite{landahl2023logicalfermionsfaulttolerantquantum}, where their encoding rates also follow an inverse-square scaling with the code distance $d$. It was mentioned in Table 1 of Ref.~\cite{algaba_fermion-qubit_2025} that the distances of the Verstraete-Cirac encoding~\cite{verstraete2005mapping}, the hexagonal encoding~\cite{chien2022optimizing,iv2024low}
and the compact encoding~\cite{derby2021compact} can all be arbitrarily extended in 2D. The extension of these codes is based on embedding these small-distance codes into surface codes with pairs of twist defects that host Majorana operators~\cite{bombin2010,brown2017}, followed by elongating the defect lines to $\Theta(d)$ and increasing the separation between different pairs of defects to $\Theta(d)$. Hence, the encoding rates scale as $\Theta(1/d^2)$.

The method of elongating defect lines has some resemblance to a concatenation of a Kitaev chain with a small-distance code~\cite{chen_private}. The logical Majorana operators are located at the endpoints of each Kitaev chain, and the pairs of Majoranas in the bulk are mapped to stabilizers. Logical operators must connect two logical Majorana operators, either on the same chain or on different chains. By increasing the lengths and separations of the Kitaev chains, the logical Majorana operators become more spatially separated from one another. As a result, the weights of logical operators connecting them increase, leading to an increase in code distance.

Our approach in 2D is different in the sense that we increase the code distance by \textit{enlarging} logical Majorana operators, meaning we encode a logical fermion into more physical fermions via color codes, instead of pulling the logical fermions apart using the Kitaev chain. In other words, for a chain of physical Majorana operators from site $1$ to site $m$, the Kitaev chain encodes logical Majorana operators on this fermionic site as $\gamma^L = \gamma_1$ and $\tilde\gamma^L = \tilde\gamma_m$ by defining $\gamma_j\tilde\gamma_{j+1}$ on all sites $j$ as stabilizers. In contrast, we treat site $1$ through site $m$ as one of the three edges of a 2D color code, and define the logical Majorana operators as $\gamma^L = \prod_{j=1}^{m}\gamma_j$, $\tilde\gamma^L = \prod_{j=1}^{m}\tilde\gamma_j$, as shown in Eqs.~(\ref{eq:logcol1}) and (\ref{eq:logcol2}). While both approaches yield the same encoding rate scaling in 2D, they lead to different scaling behaviors in 3D. 

In 3D, our encoding rate still scales as $\Theta(1/d_{Fq}^2)$ because stacking 2D color code layers next to each other, as shown in Fig.~\ref{fig:3dsummary}(c), does not enable the existence of low-weight logical operators. Since each logical Majorana operator is composed of multiple physical Majorana operators in this approach, the weight of the logical hopping back operator remains on the same order as that of other logical operators. On the other hand, the Kitaev chain approach requires each logical Majorana operator to be separated from all others by a distance of $\Omega(d)$, resulting in an encoding rate of $\Omega(1/d^3)$ in 3D. Therefore, our color code approach achieves better encoding rate scaling than the Kitaev chain approach in 3D.

It is worth noting that we expect using fermionic codes inspired by higher-rate qubit codes could improve the 3D encoding rate, as our current stacked 2D color code construction does not saturate the 3D Bravyi-Poulin-Terhal (BPT) bound~\cite{Bravyi_2009,Bravyi_2010,Flammia_2017}.

In addition, Ref.~\cite{setia2019superfast} proposed a fermion-to-qubit mapping defined on graphs with only even-degree vertices, where the encoding rate scales linearly with the graph’s maximum vertex degree. Although they proved that their mapping is robust to any single-qubit error, the relationship between the code distance and the vertex degree remains unclear. In Appendix D of Ref.~\cite{setia2019superfast}, they showed how increasing the vertex degree can result in higher-weight logical operator generators of the fermionic algebra. However, the scaling of stabilizer weights and distance depends heavily on the graph construction and the numbering (or labeling) of the edges connected to each vertex. While it may still be possible to find a construction with low stabilizer weights and high distance, we can also identify constructions where the stabilizer weights grow with the vertex degree or where the product of a few high-weight logical operator generators yields a low-weight logical operator. Constructing high-distance, low-stabilizer-weight codes with good encoding rates based on their approach is left for future investigation.

\subsection{Tradeoff for large code distance}
Lastly, we note that efficient, fault‑tolerant protocols for implementing the logical hopping and occupation operators have not yet been developed. While the large weight of logical operators helps suppress errors during storage (i.e., when the code functions as a memory), it nevertheless demands large circuits for implementing simulations. Without fault tolerance, these larger circuits are more vulnerable to errors, thereby reducing fidelity. Consequently, there may be an optimal code distance for simulation when fault-tolerant protocols are unavailable.

Furthermore, it has been pointed out~\cite{setia2019superfast} that since near-term variational quantum eigensolver experiments are more prone to errors when the logical operator weight is large~\cite{peruzzo2014variational, kandala2017hardware}, simply increasing the logical operator weight can impair performance in simulating such experiments. However, the net effect in the presence of error correction remains unclear.

\begin{acknowledgments}
We thank Yu-An Chen, Alexander Schuckert, Dong Yuan, Chao Yin, Martin Lebrat, Michael Perlin, Xiaozhen Fu, and Da-Chuan Lu for valuable discussion. We acknowledge funding from the U.S. Department of Energy, Quantum Systems Accelerator and NSF PFC grant No.\ PHYS 2317149.
\end{acknowledgments}


\bibliography{biblio}

\appendix
\begin{figure*}[t]
    \centering
    \includegraphics[width=\textwidth]{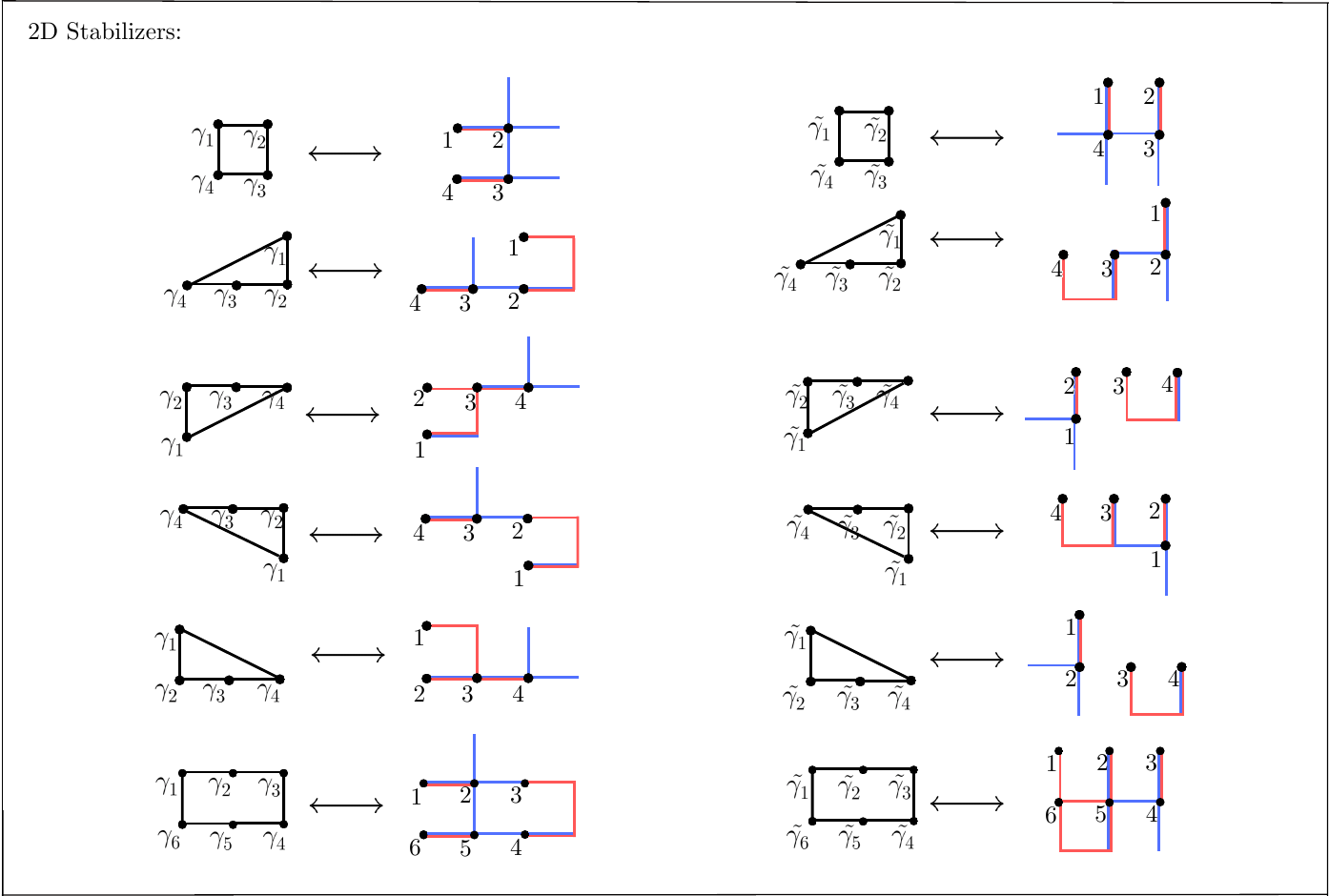}
    \caption{A complete list of all distinct 2D plaquette-type stabilizer generators from deforming the 2D color code, derived from the mapping in Eqs.~(\ref{eq:hoppingup}--\ref{eq:wa_operator}). In addition, the weight of some stabilizers has been lowered by applying stabilizer $G$.}
    \label{2d list}
\end{figure*}

\begin{figure*}[t]
    \centering
    \includegraphics[width=\textwidth]{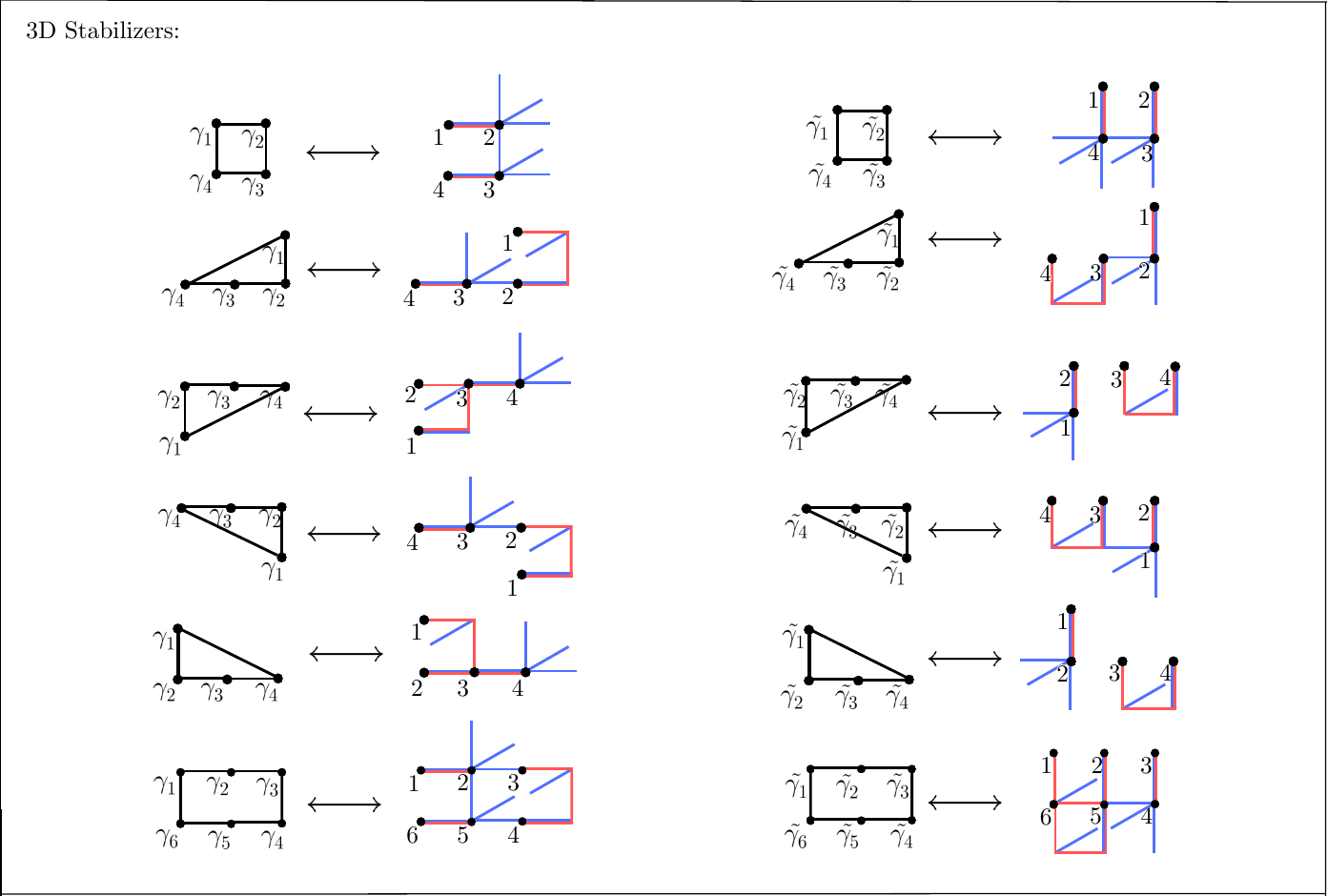}
    \caption{A complete list of all distinct plaquette-type stabilizer generators in our 3D code, derived from combining the mapping in Fig.~\ref{fig:3dmapping} with the deformed 2D color code blocks in Fig~\ref{fig:code-blocks}.}
    \label{3d list}
\end{figure*}
\FloatBarrier

\section{Complete list of plaquette-type stabilizer generators for 2D and 3D codes}\label{apx:stabilizers} 
In Fig.~\ref{fig:code-blocks}, we introduce plaquette-type stabilizers that detect errors in the fermionic algebra by deforming 2D color code blocks that encode logical Majorana modes ($\gamma^L$ and $\tilde{\gamma}^L$) onto a square lattice. The deformation of these code blocks in even and odd columns yields six distinct stabilizer generators in the Pauli algebra as shown in Fig.~\ref{2d list} and Fig.~\ref{3d list} for our 2D and 3D codes, respectively. The shapes of the stabilizers in Fig.~\ref{2d list} and Fig.~\ref{3d list} match the shapes of those in the 2D color code as shown in Fig.~\ref{fig:code-blocks} and indicate the corresponding location of each stabilizer after being mapped to Pauli algebra.

Furthermore, these stabilizers were derived through an analogous computational process to that in Eqs.~(\ref{eq:stabalizerexample}) and~(\ref{eq:stabalizerexample2}).

\section{Nontrivial Pauli $Z$ loops and membranes}\label{apx:z}

\begin{figure*}[htbp]
    \centering
    \includegraphics[width=\textwidth]{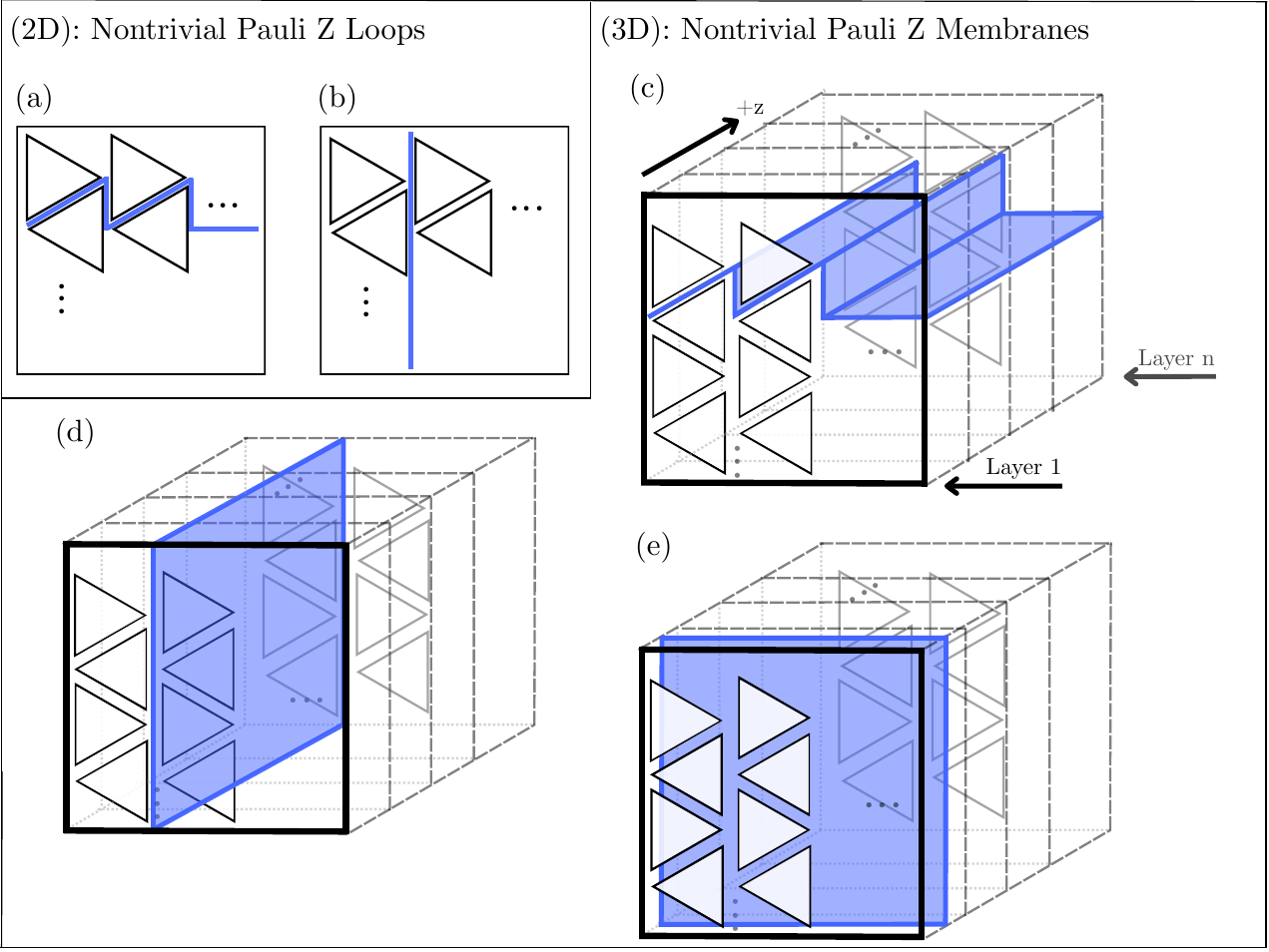}
      \caption{Nontrivial loops and membranes of Pauli $Z$ operators, both of which commute with all stabilizers. They act as generators for logical operators that change the sectors. (a)(b) Nontrivial $Z$ loops wrapping around the color code blocks without intersecting them. (c)(d) The projection of a nontrivial $Z$ membrane onto the $x$-$y$ plane matches the loop structures shown in (a) and (b). (e) A nontrivial $Z$ membrane lies vertically between adjacent layers of color code blocks.}
    \label{fig:nontrivialz}
\end{figure*}

As we mentioned before in Sec.~\ref{sec:chen2D}, all nontrivial Pauli $Z$ loops across the entire square lattice will commute with the stabilizers $G_v$ at all vertices $v$ and the padding stabilizers $W_p$. However, only a subset of them will commute with all the plaquette-type stabilizers and become logical operators/errors. As shown in Fig.~\ref{fig:nontrivialz}, when the nontrivial $Z$-loop does not intersect any color code blocks, it commutes with all stabilizers because all the Pauli $X$ components of the plaquette-type stabilizers lie within the color code blocks, as shown in Fig.~\ref{2d list}. However, if the $Z$ loop passes through a color code block, it commutes with all plaquette-type stabilizers only if it differs from the previous configuration either by logical occupation operators or by a trivial loop of Pauli $Z$ operators formed from a product of plaquette-type stabilizers. Therefore, a nontrivial $Z$-loop across the lattice that does not intersect any color code blocks, shown in Fig.~\ref{fig:nontrivialz}, is a generator of logical operators/errors of this type.

Those nontrivial $Z$ loops which serve as logical operators, denoted $Z_{NL}$, change the \textit{sector} of a state. In 2D, the fermionic vacuum state, defined as the simultaneous $+1$ eigenstate of all stabilizers and physical occupation operators, can be fixed by choosing any ground state of the 2D toric code. Fixing different fermionic vacuum states corresponds to choosing different sectors. One can show that the expectation value of any fermionic observable is unaffected by the initial choice of fermionic vacuum state, unless the observable itself also forms nontrivial loops. However, $Z_{NL}$ can affect the expectation values of observables and act as logical errors in the system if they change the sector during the simulation.

Nevertheless, $Z_{NL}$ can be neglected in practice, as its weight scales with the lattice side length, which is $\Theta(\sqrt{N_F} d_{Ff})$, where $N_F$ denotes the number of logical fermionic sites in the system. Under an adversarial error model, this error weight exceeds the weights of our logical hopping and occupation operators. Since the code distance is defined by the minimum weight of logical operators, $Z_{NL}$ does not affect the code distance. In contrast, under a stochastic noise model, the probability of such long nontrivial loops is exponentially suppressed as the number of qubits increases, either through increasing the code distance or the number of fermionic sites. Therefore, as long as the qubit number is sufficiently large, these logical errors are suppressed and do not pose a significant obstacle to quantum simulation. 

In our 3D code, the nontrivial $Z$ loops are replaced by nontrivial membranes of Pauli $Z$ operators, as illustrated in Fig.~\ref{fig:nontrivialz}. 

When the membrane is perpendicular to the 2D color code layers, the commutation and anticommutation relations are essentially the same as in the 2D case. Specifically, if the projection of the nontrivial membrane onto a 2D plane follows the same path as a nontrivial loop in the 2D code and does not intersect any color code blocks, it will commute with all plaquette-type stabilizers in the 3D code. This follows from the fact that the projected stabilizers of the 3D code onto a 2D layer are identical to the stabilizers of the 2D code, as discussed in Appendix~\ref{apx:projection}. When the membrane is parallel to the layers, the generator of this type of logical error must lie between two layers. This is because all the $X$-components of the plaquette-type stabilizers are located on the layers, while the stabilizers between layers consist only of Pauli $Z$ operators, which commute with the membrane.

These nontrivial membranes also have extensive weight, scaling with the size of one of the horizontal or vertical cross-sections of the cube. Therefore, they can also be neglected in practice.

\section{Relation between stabilizers and logical operators in 2D and 3D} \label{apx:projection}
One can observe that if we project the stabilizers and logical operators $T^L$ and $W^L$ of our 3D code to the $x$-$y$ plane, which is equivalent to removing all edges in the $z$ direction, we will exactly recover the stabilizers and logical operators of our 2D code up to applying $G_v$. This is because projecting the 3D stabilizers and physical hopping and occupation operators $T$ and $W$ in Fig.~\ref{fig:3dmapping} onto any of the $x$-$y$, $x$-$z$, or $y$-$z$ planes recovers the corresponding stabilizers and operators of the 2D code.

\section{Decoder threshold argument}\label{sec:decoder_details}

We argue that the decoder of our code, briefly mentioned in Sec.~\ref{sec:decoding}, has a threshold such that the logical error rate is exponentially suppressed with increasing code distance $d_{Ff}$ when the Pauli error rate is below the threshold. Furthermore, increasing the number of logical fermionic sites $N_F$ does not destroy this threshold, provided that $d_{Ff}$ is at least on the order of $\log(N_F)$.

The decoder is applicable to our 3D code, as well as to our 2D code when the physical-fermion-to-qubit encoding in our protocol is replaced with the $d_{fq} = 3$ mapping from Ref.~\cite{chen_error-correcting_2024}. The choice $d_{fq} = 3$ is motivated by the ability to construct a complete lookup table of single-qubit errors and their corresponding distinct syndromes.

The reasoning is as follows: Ref.~\cite{chen_error-correcting_2024} provides an explicit bijection between single-qubit errors and their corresponding syndromes for the $d_{fq} = 3$ mapping, and a similar correspondence can be derived for the 3D $d_{fq} = 3$ encoding. To decode, we first use this correspondence to map qubit errors to Majorana errors. While single-qubit errors can be corrected, the correspondence between syndromes and multi-qubit errors, such as those affecting two or more nearby qubits, is not bijective. In this case, we apply the correction operation corresponding to the lowest Pauli-weight error configuration consistent with the observed syndrome. Other possible error configurations, when combined with this correction, may correspond to physical hopping or occupation operators, thereby resulting in Majorana errors $\gamma$ and $\tilde{\gamma}$. Majorana errors can be correlated due to this mapping; however, these correlations are short-range and decay exponentially with the spatial separation between errors. This exponential decay arises because generating a pair of spatially separated correlated Majorana errors requires a Pauli error whose weight increases at least linearly with the spatial separation between the Majorana errors. Therefore, we expect that a threshold still exists when we treat $\gamma$ and $\tilde{\gamma}$ at each site as independent errors, each occurring with probability $P$, despite the presence of correlations ~\cite{Aharonov2006,Chubb_2021}.

If either a $\gamma$ or a $\tilde{\gamma}$ error falls on the padding region, the error will anticommute with the padding stabilizer $W_p$ and produce the same syndrome. However, the error can be easily corrected in either case by applying a $\gamma$ operator, as it either cancels the error via $\gamma_p^2 = 1$ or combines with it to form a stabilizer, since $\gamma_p \tilde{\gamma}_p \propto W_p$.

The $\gamma$ and $\tilde{\gamma}$ errors within a fermionic color code can be treated separately, analogous to how Pauli $X$ and $Z$ errors are handled independently in a qubit color code. A string of $\gamma$ (or $\tilde{\gamma}$) errors either violates the two plaquette-type stabilizers at its endpoints that have $\gamma$ (or $\tilde{\gamma}$) operators on all their vertices, or terminates at one of the sides on the boundary of the color code block. Therefore, the decoders with thresholds for 2D qubit color codes~\cite{sarvepalli2012efficient, kubica2023efficient, lee2025color} can be directly adapted to decode the fermionic color code. Each color code block can be decoded independently using a decoder with threshold $P_{th}$.

When $P < P_{th}$, the logical error rate is expected to be exponentially suppressed with increasing code distance, meaning that the probability of a logical error $\gamma^L$ on a single color code block, which is equal to the probability of a $\tilde{\gamma}^L$ error, is given by
\begin{equation}
P_b \propto e^{-\alpha d_{Ff}},
\end{equation}
where $\alpha$ is a positive constant~\cite{fowler2012,Takada_2024,google2024}. 

Based on the previous argument that $\gamma$ and $\tilde{\gamma}$ errors can be treated independently, the probability that a single code block is decoded successfully is $(1 - P_b)^2$, which corresponds to the probability of having neither a $\gamma^L$ nor a $\tilde{\gamma}^L$ error on that code block. 

When there are $N_F$ code blocks, the probability that no logical error occurs on any of them is given by
\begin{equation}
1 - P_L = (1 - P_b)^{2N_F} \approx 1 - 2N_F P_b,
\end{equation}
where $P_L$ is the logical failure rate of the entire code, and the approximation in the last step is obtained using the binomial expansion. To keep $P_L$ constant as $N_F$ increases, we require
\begin{equation}
2N_F e^{-\alpha d_{Ff}} \sim \Theta(1),
\end{equation}
and therefore $d_{Ff} \sim \log N_F$.

\end{document}